# Generic Cosmic Censorship Violation in anti de Sitter Space

Thomas Hertog[1], Gary T. Horowitz[1], and Kengo Maeda[2]

[1] Department of Physics, UCSB, Santa Barbara, CA 93106
[2] Yukawa Institute for Theoretical Physics, Kyoto University, Kyoto 606-8502, Japan

We consider (four dimensional) gravity coupled to a scalar field with potential $V(\phi)$. The potential satisfies the positive energy theorem for solutions that asymptotically tend to a negative local minimum. We show that for a large class of such potentials, there is an open set of smooth initial data that evolve to naked singularities. Hence cosmic censorship does not hold for certain reasonable matter theories in asymptotically anti de Sitter spacetimes. The asymptotically flat case is more subtle. We suspect that potentials with a local Minkowski minimum may similarly lead to violations of cosmic censorship in asymptotically flat spacetimes, but we do not have definite results.

Perhaps the most important open problem in classical general relativity is to prove (or find a counterexample to) Penrose's cosmic censorship hypothesis [1]. This states that physically reasonable initial data cannot produce naked singularities, i.e., singularities that are visible to distant observers. Despite the fact that our current theory of black holes is heavily based on this hypothesis, there is rather little direct evidence that it is true.

We will show that (weak) cosmic censorship can be violated rather easily when gravity is coupled to a scalar field in asymptotically anti de Sitter (AdS) spacetimes. While it is known that naked singularities can be produced with pressureless matter (see, e.g. [2]) or fine tuned initial conditions [3], this is one of the first examples of a generic violation of cosmic censorship for "reasonable" matter (see [4] for an asymptotically de Sitter example). In addition to restricting the nature of a possible cosmic censorship theorem applicable to our universe, this result may be of interest in string theory. It has been argued that string theory with AdS boundary conditions is completely described by a conformal field theory [5]. While we have not attempted to derive our action from string theory, if naked singularities also arise in that case, one could probably use the dual field theory description to see how they are resolved in a quantum theory of gravity.

We consider four dimensional gravity, coupled to a single scalar field with a potential $V(\phi)$. We take $V$ to have a global minimum at $\phi = 0$ and a local minimum at $\phi = \phi_1 > 0$ (see Fig. 1). We assume $V(0) = -3V_0$ and $V(\phi_1) = -3V_1$ are both negative and we consider solutions that asymptotically approach the local (AdS) minimum at $\phi_1$. We require that $V$ satisfies the positive energy theorem (PET) for solutions with this boundary condition. While some formulations of this theorem assume a local energy condition stating that $V$ is never less than its asymptotic value, it has been shown that this is not necessary [6, 7]. Generally speaking, the PET holds if the barrier separating the extrema is high enough, but it does not hold if the barrier is too low [10]. By adjusting the height of the barrier to be close to the transition point, one decreases the mass of nontrivial configurations that probe the region of $V$ around the true minimum. We will show that although the positive energy theorem holds in such theories, cosmic censorship does not. We demonstrate this by first constructing initial data with a large approximately homogeneous region in the interior where $V < -3V_1$, but with $\phi \to \phi_1$ asymptotically. The central region evolves to a singularity, since a homogeneous scalar field rolling down a potential to a negative minimum will generically become singular. We then show that if the barrier is close to the transition point, the total mass is too small to produce a black hole large enough to enclose the entire singular region, so the singularity must be naked.

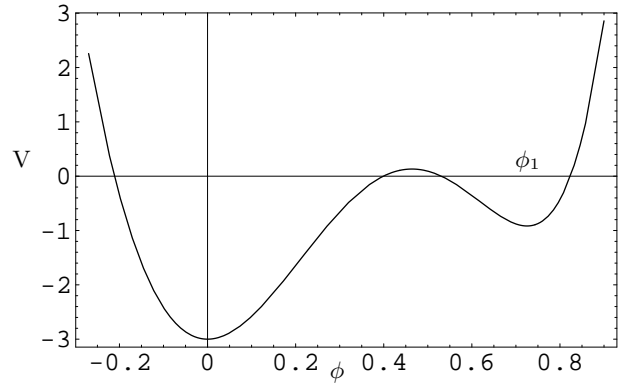

FIG. 1: A potential $V(\phi)$ that satisfies the positive energy theorem for solutions that asymptotically approach the local (AdS) minimum at $\phi_1$, but which violates cosmic censorship.

This violation of cosmic censorship in AdS is quite general since for a large class of potentials, one only has to adjust one parameter. Even though the naked singularity in black hole critical phenomena [3] also arises from adjusting one parameter, the implication here is completely different. This is because we are adjusting a parameter in the potential, not the initial data. For a given theory, there is an open set of initial data which produce naked singularities. Furthermore, in our case one does not even have to fix the parameter exactly, it only has to be close to some critical value.

It may also be possible to violate cosmic censorship for asymptotically flat initial data, using potentials of the above form with the local minimum at $V = 0$. However it is much easier in the asymptotically AdS case. This is because a large black hole of radius $R_s$ in AdS requires a mass $M_{BH} = (R_s^3 + R_s)/2$ (where we have set the AdS radius to one). This is much larger than the mass of a Schwarzschild black hole of size $R_s$. For this reason, the asymptotically flat case is much more delicate. We will discuss it at the end, but not come to a definite conclusion. We will see that this can be explored with 1+1 dimensional numerical relativity and hence provides a feasible new test of cosmic censorship.

To begin, we find the precise condition for potentials of the above type to admit a PET. To minimize the mass, we consider initial data with all time derivatives set to zero. For time symmetric initial data the constraint equations reduce to

$$^{(3)}\mathcal{R} = g^{ij}\phi_{,i}\phi_{,j} + 2V(\phi) \tag{1}$$

where we set $8\pi G = 1$. Since spatial gradients raise the energy, we first restrict attention to spherically symmetric configurations with metric

$$ds^2 = \left(1 - \frac{m(r)}{4\pi r}\right)^{-1} dr^2 + r^2 d\Omega_2^2. \tag{2}$$

The constraint then yields the following equation for the "mass" $m$ as a function of the radius,

$$m_{,r} + \frac{1}{2}mr(\phi_{,r})^2 = 4\pi r^2 \left[V(\phi) + \frac{1}{2}(\phi_{,r})^2\right] \tag{3}$$

The general solution for arbitrary $\phi(r)$ is

$$m(r) = 4\pi \int_0^r e^{-\int_{\tilde{r}}^r \hat{r}(\phi_{,r})^2/2\, d\hat{r}} \left[V(\phi) + \frac{1}{2}(\phi_{,r})^2\right] \tilde{r}^2 d\tilde{r}. \tag{4}$$

The total ADM mass is defined to be

$$M = \lim_{r\to\infty} [m(r) + 4\pi V_1 r^3] \tag{5}$$

We require that $\phi \to \phi_1$ faster than $1/r^{3/2}$ since this is required for finite mass. In fact, it suffices to consider configurations where $\phi(r)$ reaches $\phi_1$ at a (possibly large) finite radius $r = R_1$, and in this case $M = m(R_1) + 4\pi V_1 R_1^3$. This is because one can always perturb $\phi$ to have finite $R_1$, keeping the change in the mass arbitrarily small.

To identify the criterion on $V$ for the PET to hold, we first minimize

$$m_V = 4\pi \int_0^{R_1} e^{-\int_r^{R_1} \hat{r}(\phi_{,r})^2/2\, d\hat{r}} V(\phi) r^2 dr \tag{6}$$

over a suitable class of $\phi$. Introducing a new radial variable $y = r/R_1$ and writing $\tilde{\phi}(y) = \phi(R_1 y)$ it is easy to see that $m_V/R_1^3$ is independent of $R_1$. Let $\mathcal{S}$ be the set of all $\tilde{\phi}(y)$ with $\tilde{\phi}(0) = \phi_0 \geq 0$, $V(\phi_0) < -3V_1$ and $\tilde{\phi}(1) = \phi_1$. The boundary condition at the origin is chosen so that

if $V$ admits any negative energy configurations, then it admits some in $\mathcal{S}$. We define

$$\rho_V \equiv \min_{\tilde{\phi} \in \mathcal{S}} \frac{m_V}{4\pi R_1^3} = \min_{\tilde{\phi} \in \mathcal{S}} \int_0^1 e^{-\int_y^1 d\hat{y}\hat{y}\tilde{\phi}'^2/2} Vy^2 dy \tag{7}$$

where $\tilde{\phi}' = \tilde{\phi}_{,y}$. The minimum clearly exists since the integral is bounded below by $-V_0$. Clearly $\rho_V$ is a continuous function of $V$, and $R_1$ is now a free parameter that acts like an overall scale. If $\rho_V < -V_1$ then the PET does not hold, since the contribution to the mass from the $(\phi_{,r})^2$ term is proportional to $R_1$ while the contribution from $V$ is proportional to $R_1^3$. So for large $R_1$, the mass will be negative. However, if $\rho_V \geq -V_1$, then the PET will hold since this minimal configuration has positive mass. (When the PET holds, the true minimal configuration has zero mass and corresponds to $\phi(r) = \phi_1$ for all $r$. Our minimal configuration has positive mass, since it is required to have $V(\phi(0)) < -3V_1$.)

To compute $\rho_V$ for a given theory we take the variation $\delta\phi$ of the integral (7), to find the lowest mass configuration subject to the boundary conditions discussed above. This yields the following integro-differential equation for the 'optimal' paths $\tilde{\phi}(y)$,

$$\int_0^{\bar{y}} dy\, y^2 V(\tilde{\phi}) e^{-\int_y^1 d\hat{y}\, \hat{y}\tilde{\phi}'^2/2} = \tag{8}$$

$$-\frac{e^{-\int_y^1 d\hat{y}\, \hat{y}\tilde{\phi}'^2/2} \left(\bar{y}^2 V_{,\tilde{\phi}} + \bar{y}^3 \tilde{\phi}' V(\tilde{\phi})\right)}{\bar{y}\tilde{\phi}'' + \tilde{\phi}'}$$

where all derivatives on the right hand side are evaluated at $\bar{y}$. Notice that the left hand side is precisely $m_V(\bar{y})/4\pi R_1^3$, so equation (8) expresses the density $\rho_V$ of the optimal paths in terms of field derivatives at $y = 1$. Subtracting the cosmological constant term $-V_1$ gives

$$\rho_V + V_1 = \frac{V_1 \left(4\tilde{\phi}' + \tilde{\phi}''\right)}{\tilde{\phi}' + \tilde{\phi}''}, \tag{9}$$

which yields yet another way to state the precise condition for the PET to hold: for the potential $V$ at the transition point, the lowest mass configuration within the class $\mathcal{S}$ has $\tilde{\phi}'' = -4\tilde{\phi}'$.

To give a concrete example, we numerically solve eq. (8) and compute $\rho_V$ for the following one-parameter family of potentials (shown in Fig. 1 for $\alpha = 45.9$),

$$V(\phi) = -3 + 50\phi^2 - 81\phi^3 + \alpha\phi^6. \tag{10}$$

We have chosen the parameter $\alpha$ to control the height of the barrier between both extrema. For $\alpha = 45.928$ we have $\rho_V = -V_1$. For this potential, $V_0 = 1$, $V_1 = .305$, and $\phi_1 = .725$. The solution for the optimal path in the theory at the transition point is shown in Fig. 2. The solution starts at the global minimum at the origin $y = 0$, climbs very slowly out the true vacuum and reaches the false vacuum at $y = 1$.

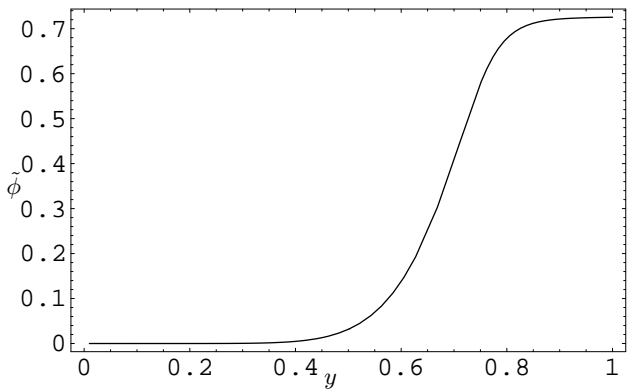

FIG. 2: The lowest mass configuration $\tilde{\phi}(y)$, subject to the boundary conditions discussed in the text, for the potential shown in figure 1.

At the transition point, $\rho_V + V_1 = 0$, the potential contribution to the mass vanishes. In terms of the area coordinate $r = yR_1$, the total ADM mass of the minimal configuration $\phi(r)$ is then given by

$$M = 2\pi \int_0^{R_1} e^{-\int_r^{R_1} \hat{r}(\phi_{,r})^2/2 \, d\hat{r}} (\phi_{,r})^2 r^2 dr \propto R_1 \ . \quad (11)$$

We now show that this lowest mass configuration evolves to form naked singularities. In the central region, the field $\tilde{\phi}(y)$ changes slowly. Take $y = y_0$ to be the radius where $m(y_0)$ deviates by one percent from its value in pure AdS, with $\Lambda = -3V_0$ (in our example, $y_0 = .58$). For large $R_1$, this corresponds to a proper radial distance

$$L \approx \int_0^{y_0 R_1} \frac{dr}{[1 + (V_0 r^2)]^{1/2}} \approx V_0^{-1/2} \ln R_1 \ . \quad (12)$$

Hence there is a large region of approximately constant density and we can model the evolution inside its domain of dependence by a $k = -1$ Robertson-Walker universe: $ds^2 = -dt^2 + a^2(t)d\sigma^2$ (where $d\sigma^2$ is the metric on the unit hyperboloid). The field equations are

$$\frac{\ddot{a}}{a} = \frac{1}{3}[V(\phi) - \dot{\phi}^2] \quad (13)$$

$$\ddot{\phi} + \frac{3\dot{a}}{a}\dot{\phi} + V_{,\phi} = 0 \quad (14)$$

and the constraint equation is

$$\dot{a}^2 - \frac{a^2}{3}\left[\frac{1}{2}\dot{\phi}^2 + V(\phi)\right] = 1 \ . \quad (15)$$

If the scalar field was exactly at the minimum of the potential, it would remain constant, and the solution would be AdS with $a(t) \propto \cos\sqrt{V_0}t$. In this case, $a = 0$ is just a coordinate singularity. The fact that the scalar field starts slightly above the minimum has a dramatic consequence. Now $\phi$ starts to roll down the potential, and when $a \to 0$, this results in a curvature singularity as we now show.

Near the minimum of the potential, $V = -3V_0 + m^2\phi^2/2$. We start with $\phi = \epsilon$, so initially we have $\phi(t) = \epsilon \cos mt$ and $a(t) = V_0^{-1/2}\cos\sqrt{V_0}t$. Define $E \equiv \dot{\phi}^2/2 + m^2\phi^2/2$. Then from (14), $\dot{E} = \dot{\phi}(\ddot{\phi} + m^2\phi) = -3(\dot{a}/a)\dot{\phi}^2$. Since $m \gg \sqrt{V_0}$, $\phi$ will oscillate many times before $a$ changes. Averaging over one period yields $\langle\dot{E}\rangle = -3(\dot{a}/a)\langle E\rangle$ which implies $\langle E\rangle a^3$ remains constant. (This could also be seen from the fact that the pressure $p = \dot{\phi}^2/2 - m^2\phi^2/2$ averages to zero.) Putting in the initial value of $E$ yields $\langle E\rangle = \epsilon^2 m^2/2a^3$. Substituting into the constraint (15) yields

$$\dot{a}^2 + V_0 a^2 - \frac{\epsilon^2 m^2}{6a} = 1 \quad (16)$$

This shows that when $\epsilon^2 m^2/a \sim 1$, $a(t)$ changes from $\cos\sqrt{V_0}t$ to $a(t) \propto (T_s - t)^{2/3}$. Near $t = T_s$, it follows from (14) that $\phi$ diverges, producing a curvature singularity, and $a(t)$ changes to $a(t) \propto (T_s - t)^{1/3}$. Since this homogeneous evolution produces trapped surfaces, the slightly inhomogeneous collapse of our central region will also produce singularities.

If this singularity lies inside a black hole, then we can trace the null geodesic generators of the event horizon back to the initial surface, where they will form a sphere of radius $R_s$. The area theorem for black holes only requires the null convergence condition and hence still holds even in theories with $V(\phi) < 0$. Since the area of the event horizon cannot decrease during evolution and the mass cannot increase, the initial mass $M$ must be large enough to support a static black hole of size $R_s$. Since a large Schwarzschild AdS black hole requires a mass $M_{BH} \propto R_s^3$ and the initial data only have mass $M \propto R_1$ this is clearly impossible if $R_s \propto R_1$, which we will soon establish. Potentials of the type we are considering admit black hole solutions with scalar hair [8], but they also have a mass which grows like $R_s^3$ if the PET holds, so they cannot be produced in the evolution.

To compute $R_s$, define $2l$ to be the proper distance on the initial surface between the boundary of the homogeneous region at $r = y_0 R_1$ and an inner radius ($R_s$) such that an outgoing radial null geodesic from the inner radius meets an ingoing radial null geodesic from $r = y_0 R_1$ at the singularity. From the Robertson-Walker form of the metric, $l = a(0)\int_0^{T_s} dt/a(t)$. From the above analysis, $l$ depends on $y_0$ through $\epsilon$, but is independent of $R_1$. (In our example, $l = 2.4$ and $T_s = 1.3$). The proper radial distance on large scales is proportional to $\ln r$, so $R_s = y_0 R_1 e^{-2l}$. Since $l$ is independent of $R_1$, this shows that $R_s$ indeed scales linearly with $R_1$. The mass grows linearly with $R_1$, but the black hole of this size requires a mass which grows like $R_1^3$. Clearly, for large $R_1$ there is not enough mass to form a black hole which encloses the singular region.

Inside the domain of dependence of the central region of radius $y_0 R_1$, the singularity will be spacelike, like a big crunch. The singularity is likely to extend somewhat outside the domain of dependence (so our estimate for $R_s$ is really a lower limit), but not reach infinity. The endpoint of the singularity will thus be naked. If the singularity did reach infinity, it would cut off all space, producing a disaster much worse than naked singularities. But this is unlikely since there would then be a radius $R_c$ on the initial surface such that the outgoing null surfaces for $r > R_c$ expand indefinitely and reach infinity, while those with $r < R_c$ hit the singularity and (probably) contract to a point. This indicates that the surface with $r = R_c$ would reach a finite radius asymptotically, just like the stationary horizons which are ruled out.

So far we have discussed just one particular initial data set, and argued that it evolves to a naked singularity. However since the difference between the available mass and the mass required to enclose the singularity inside a black hole is very large, it is clear that small perturbations in the geometry or adding small initial time derivatives will not change our conclusion. The evolution will still produce a naked singularity.

For this reason, our construction leads to *generic* violations of cosmic censorship. Moreover, it is not necessary to tune the potential exactly. Once one has a theory with configurations that produce naked singularities for some $R_1$, one can raise the height of the barrier slightly (increasing the mass by $\varepsilon R_1^3$) and still not have enough mass to form a black hole.

It is natural to ask if the same type of potential with a local minimum at $V(\phi_1) = 0$, will lead to violations of cosmic censorship for asymptotically flat spacetimes. This is possible since one can still satisfy the positive energy theorem even when the potential has a negative global minimum, and an (approximately) homogeneous scalar field rolling down the potential to a negative minimum again produces a singularity. However this question is more difficult to answer since a large Schwarzschild black hole only requires a mass proportional to its radius. Thus even after adjusting the height of the potential to cancel the $R_1^3$ contributions to the mass, one may still have enough mass to enclose the singularity in a black hole of size $R_s$.

Nevertheless, we believe the singularity may be naked also in this case, for the following reason. Our minimal configuration can be viewed as a region of negative energy proportional to $-R_1^3$ surrounded by a shell of positive energy proportional to $+R_1^3$, leaving the ADM energy proportional to $R_1$. Since the initial data are time symmetric, one expects some of the energy in the shell to radiate to infinity. However, even if only a small fraction of the energy in the shell is radiated away, the Bondi mass would become negative. (This is not an issue in the asymptotically AdS case if one uses reflecting boundary conditions as required by string theory.) To ensure positivity of the Bondi energy, one can increase the height of the barrier. By continuity, there must be a point where the final Bondi mass remains positive but small. We suspect that there are potentials for which the final Bondi mass of the minimal configurations is too small to produce a black hole surrounding the singular region, leading to violations of cosmic censorship for asymptotically flat spacetimes. One clearly needs to study the full evolution to explore this possibility, but since one can test this with spherically symmetric configurations, this should be an easy problem for numerical relativity.

One might argue that any matter theory that does not satisfy the dominant energy condition is unphysical. However this seems unreasonable in light of the fact that a large class of supersymmetric compactifications $M_4 \times K$ contain four dimensional potentials which become negative [9]. It would be very interesting to see if initial data similar to what we have studied here evolve to naked singularities in some supersymmetric compactifications.

The fact that cosmic censorship may not hold in our universe could be viewed as a desirable feature. If singularities can be visible, one has the possibility to vastly extend the range over which general relativity could be tested experimentally, and to directly observe effects of quantum gravity associated with high curvature.

**Acknowledgments**

It is a pleasure to thank D. Birmingham, J. Isenberg, D. Page and J. Walcher for helpful discussions. This work was supported in part by NSF grants PHY-0070895, PHY-0244764, and a Yukawa fellowship.